# SKYGLYPHS: REFLECTIONS on the DESIGN of a DELIGHTFUL VISUALIZATION

**Bon Adriel Aseniero**
Autodesk Research
661 University Avenue,
Suite 200 Toronto, ON
M5G 1M1 Canada
bon.aseniero@
autodesk.com

**George Fitzmaurice**
Autodesk Research
661 University Avenue,
Suite 200 Toronto, ON
M5G 1M1 Canada
george.fitzmaurice@
autodesk.com

**Sheelagh Carpendale**
Simon Fraser University
8888 University
Drive Burnaby, BC
V5A 1S6 Canada
sheelagh@sfu.com

**Justin Matejka**
Autodesk Research
661 University Avenue,
Suite 200 Toronto, ON
M5G 1M1 Canada
justin.matejka@
autodesk.com

**Abstract**

In creating *SkyGlyphs*, our goal was to develop a data visualization that could possibly capture people's attention and spark their curiosity to explore a dataset. This work was inspired by a *mingling* of research including serendipitous interactions, visualizations for public displays, and personal visualizations. SkyGlyphs is a nonconventional whimsical visualization, depicting datapoints as animated balloons in space. We designed it to encourage non-experts to casually browse the contents of a repository through visual interactions like linking and grouping of datapoints. Our contributions include SkyGlyphs' representation and our design reflection that reveals a perspective on how to design delightful visualizations.

## Introduction

Artistic and out of the ordinary representations of data—which we henceforth group into the term *nonconventional*—have historically appeared throughout the information visualization (Infovis) literature. Data visualization (Datavis) designers often create *nonconventional data representations* that go beyond traditional techniques like bar charts, pie charts, and scatterplots to adapt to emergent contexts that are more personal, leisurely, or communicative. Some examples include embellished charts [7], artistic visualizations [26,27], data sketches [12], and data comics [6].

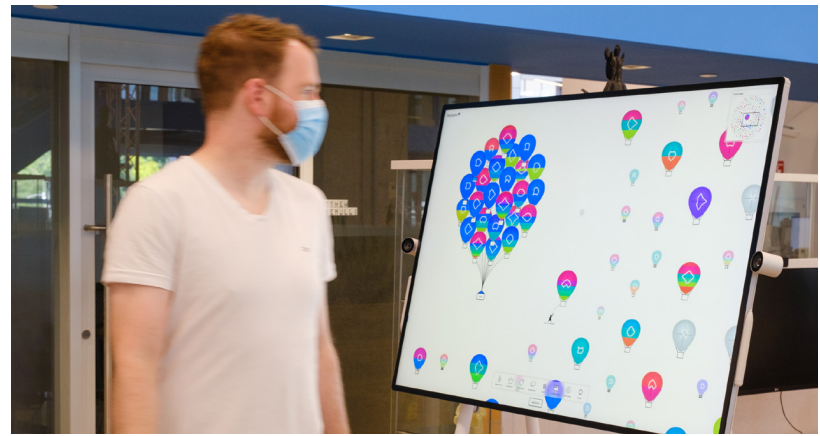

A person walks by a large screen display showing **SkyGlyphs**, a data visualization that we designed to have a nonconventional data representation to captivate people's attention and leverage their curiosity to explore a dataset.

Beyond supporting typical data analysis and exploration tasks, designers of nonconventional datavis consider the aesthetics of their datavis and anticipate how their designs fit into people's routines and environments. Like traditional datavis, nonconventional datavis can lead people to form deeper insights about their data [26]. Furthermore, designers have designed nonconventional datavis that elicit pleasant emotions through aesthetic appeal [5] and provide enjoyable experiences overall [24]. We call this subset as **"delightful visualizations."**

Nevertheless, the design of nonconventional data representations has mostly been left to the datavis designers themselves. Most of our current understanding of designing data visualizations are informed by task-oriented professional or work contexts, and we as a community have much to learn about designing nonconventional representations for emergent contexts. It is important to uncover new perspectives on how a designer can design new data representations. These perspectives can be applied to emergent research on datavis such as casual and personal visualizations, which includes visualizations for public engagement and data-driven journalism, where end-users tend to be non-experts, and where aesthetics and engagement may take precedence over task efficiency [17,23].

To take on this challenge, in this pictorial, we present a storyline of our emergent design process in creating a datavis tool—*SkyGlyphs*—

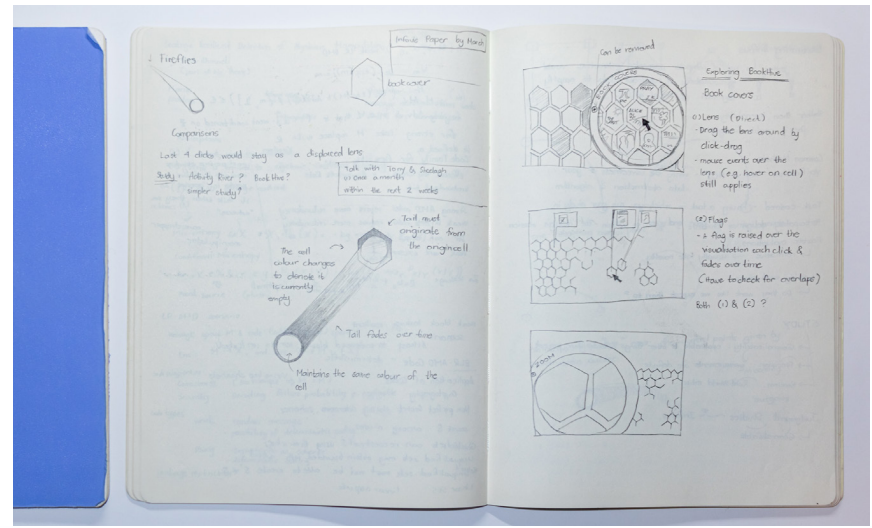

Photo of a datavis designer's sketchbooks in which they sketch their datavis designs before implementing them.

which features a delightful, nonconventional representation of datapoints as "balloons" floating in space. Through an intentional use of whimsical aesthetics including vibrant colours, elegant shapes, and subtly fun animations in our design, our aim is to captivate people's attention and leverage their curiosity as entry points to data exploration.

We applied our representation to visualize a repository of ~3,500 presentation decks (~90,000 slides) created and shared by employees within a company. We provide contextual explanation of the challenges that lead use to its design, and supplement it with sketches and illustrations to paint a more complete picture of our design process. We then provide some reflections on our design and process to discuss the potential of nonconventional representations that leverage *delight* and *curiosity* as entry points to visual data exploration.

### Design-Study: A Visualization for a Company's Data Repository

We developed SkyGlyphs as part of Autodesk's[1] corporate intelligence project concerned with utilizing the vast amount of *organic data* that employees generate and share across the company. For instance, presentation slides are often made by employees to disseminate key information that could benefit the company. However, these are underutilized often being shared once then forgotten. Drawing parallels in education, presentation slides posted online can be used by students who missed a lecture or by self-learners of the materials' topic [28].

Autodesk employees typically shared their presentation documents in a browser repository. This repository allows a person to search for a document they are looking for (targeted search) but it lacks support for a more free-form, exploratory browsing of documents. Through some form of serendipitous and incidental learning, we supposed that encouraging employees to revisit such materials might lead them to become more aware of company information (e.g., by finding content about products they are working on), encourage collaboration (finding people who share similar content), and spark new ideas. This active business community often do peruse slides for keeping information current and for possible slide re-use.

___________
1   https://autodesk.com

### Design Challenge

Design a datavis that would visualize the company's repository of presentation documents in a way that would encourage and enable employees to browse its contents in a free-form manner.

### Design Goals

We distilled several design goals from the Infovis and Human-Computer Interaction (HCI) literature to ensure that, overall, our datavis is *engaging*—an important factor in helping people gain insight [30].

#### Enable Serendipitous Browsing

One requirement we needed to address in our design is to support a free-form method of browsing the company's data repository. While direct search is best for quickly finding a specific item a person is looking for, free-form browsing is most effective for enabling a person to find new items that they may be interested in [1]. This process supports *serendipitous browsing*—from "serendipity," which is described as the accident of finding something that is of value while not specifically looking for it [25]—relies on stimulating a person's sensations that would make them curious or drawn to a certain items. For example, in *the Bohemian Bookshelf*, Thudt et al. visualized a book's cover, size, and topics using nonconventional visualization techniques that engaged people [25]. This serendipitous experience can be utilized to give people a different form of recommendation or incidental learning—one that is personal while maintaining a notion of freedom of choice.

#### Leverage Curiosity

A barrier we faced is that employees are not readily invested in exploring the company's repository, as they may have other responsibilities, or they might not be aware of it. Hence, inspired by work on visualizations on public displays, one of our goals is to leverage a person's curiosity as an entry point to visual data exploration. While not necessarily an emotional investment, curiosity is a motivator that can initiate further exploration of a dataset [2] and encourage the sustained use of a datavis [11]. One method that motivate curiosity involves introducing visual complexity and uncertainty [8,20].

#### Incite Delight

The task we needed to support involve capturing people's attention and encouraging them to browse a data repository. This has a lot in common with contexts supported by public visualizations [15] and personal visualizations, in which *enjoyability* is a factor [24]. Furthermore, we also considered previous works on embellished charts that showed how intentional visual embellishments can improve the *memorability* of data representations [7,9]. Thus, as an artistic choice, we made it our goal to incorporate whimsical aesthetics into our datavis design regardless of it being for a work-related task. As we observed from other datavis designers in practice [4,12,19,24], through the intentional use of vibrant colours, elegant shapes, and fluid animations in our design, our aim is to induce a *pleasant feeling of satisfaction*—or *delight*—from the people using or viewing them.

**Design Solution: Balloon Metaphor**

To develop our datavis solution, our design process involved balancing between functionally conveying data and expressing our more artistic, authorial intent. Hence, we went through multiple design iterations using sketches and storyboards. Our chosen design solution is a variation of a *unit visualization*, in which each datapoint is represented by a unique visual item [22].

Based on the first author's personal aesthetic choice and feedback from the rest of the team, we chose a *balloon metaphor* to represent each datapoint in our data repository. This choice was informed by a careful consideration of visual aesthetics using vibrant colours, elegant shapes, and fluid animations that could help induce delight. In developing this metaphor, we made use of intrinsic visual qualities of different kinds of balloons as integral parts of our data representation. For example, we were inspired by hot-air balloons that typically have colourful patterns and shapes, as well as parts like strings and baskets that could be abstracted. We also examined how to leverage behavioural characteristics of balloons: they can float, move with the wind, be bundled, carry away items, and *pop*.

There is also an inherent playful and amusing quality associated with balloons. In western culture, balloons are used in delightful events like celebrations and parties. Given that one of our goals is to incite delight, we decided to use this recognizable metaphor that already has positive associations.

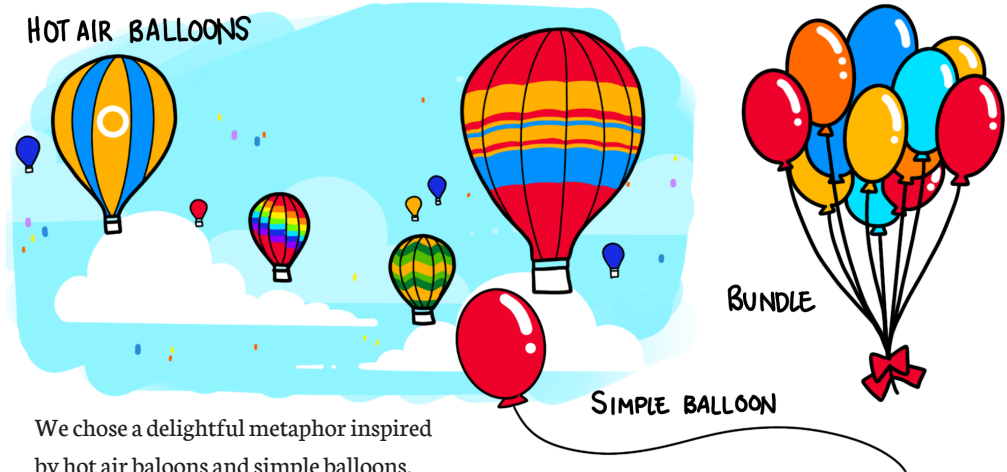

We chose a delightful metaphor inspired by hot air baloons and simple balloons.

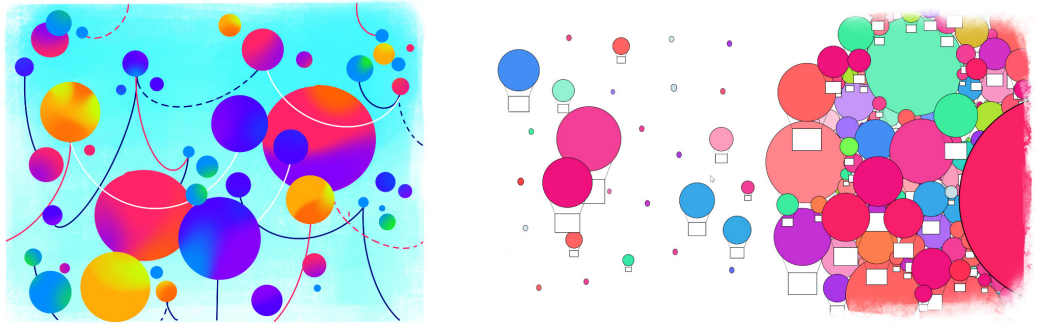

We explored different ways of how to abstract the balloon metaphor, ranging from highly abstracted to more realistic or figurative styles.

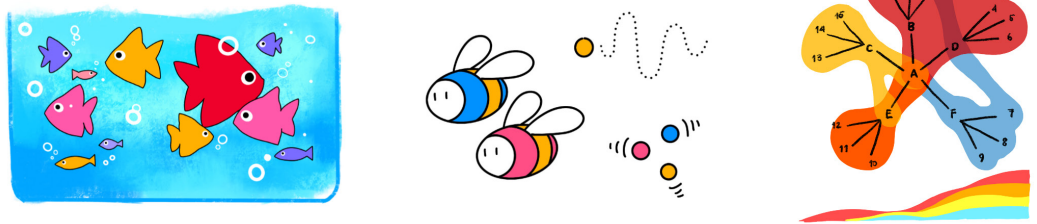

We also sketched other possible metaphors from inspirations like fish tanks, swarming insects, and combining familiar abstractions like bubble sets and stackgraphs.

The sketches in this page illustrate how we envisioned the balloon metaphors for individual slides and slide decks. To formalize the data-coding pattern of the balloons, we mapped metadata attributes of presentation documents such as who shared it and when it was shared, as well as other attributes like number of slides, words, buzzwords, and company-relevant keywords, which we extracted from the content of each document. In these sketches, we illustrate how the colour of the balloon could be mapped into categorical data attributes. Then, a glyph in the middle of the balloon could be used to depict numeric data attributes about the slide content. The balloon string could then be used to depict relationships with other datapoints.

Throughout the design process the first author held frequent co-design sessions with team members from the company with considerable business experience. Many decisions were made based on their input. For example, the four axes were chosen to include sufficient yet not overwhelming complexity. Products, company-relevant keywords and buzzwords could act as search tags. Practically speaking, while the keywords/buzzwords themselves are more important than the amount appearing in the slide deck, we wanted to upfront these attributes as teasers where one might ask *"why does this slide deck have a lot of keywords? What are these keywords?"* They can then explore the attributes in more detail. Meanwhile, the size of the slide deck might affect a viewer's decision on whether to look at it or not based on their available time.

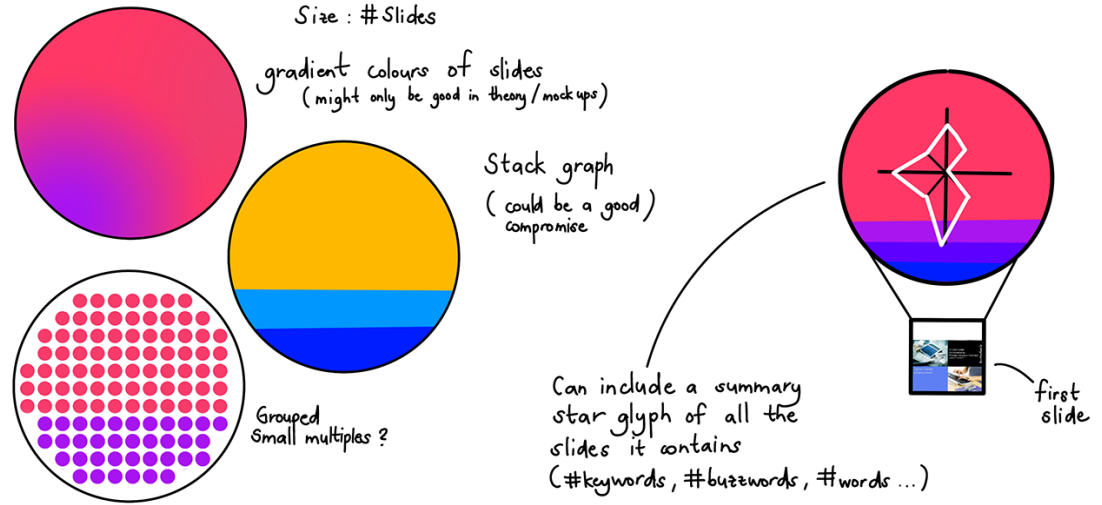

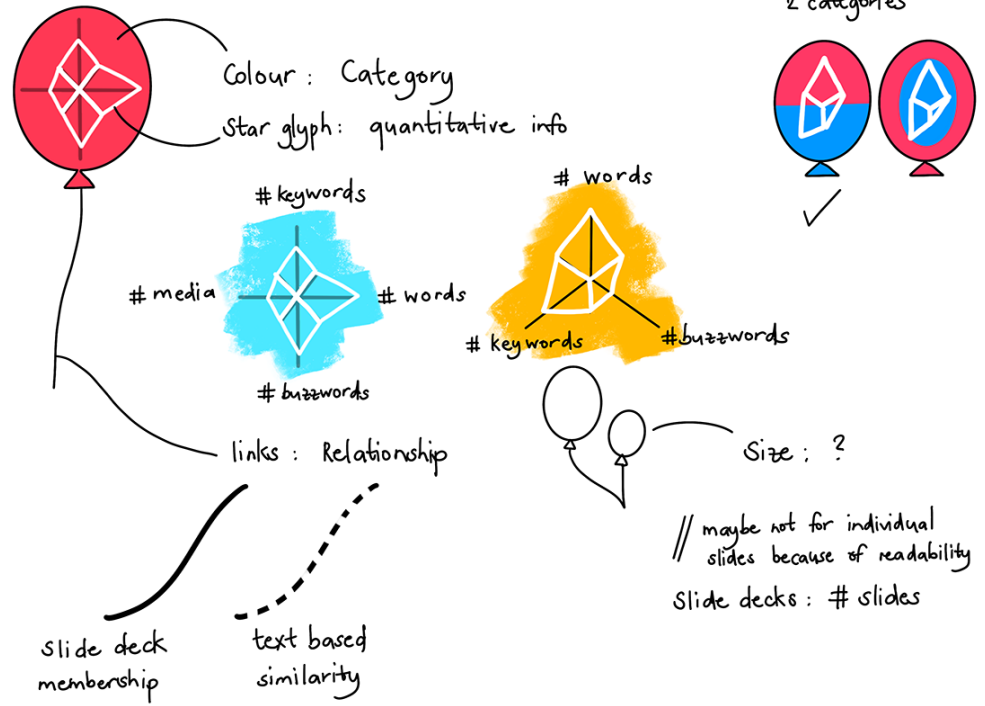

## Interaction Sketches

In addition to sketching the data representation design, we also used sketches to envision how people could interact with the datavis.

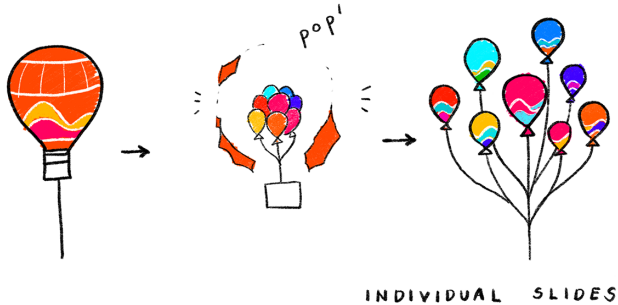

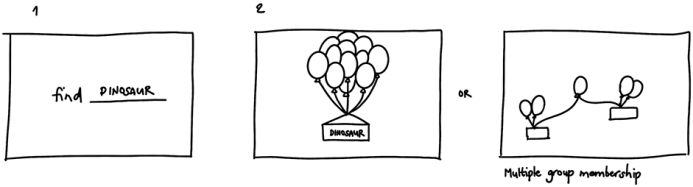

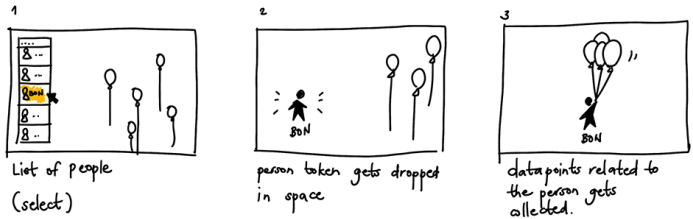

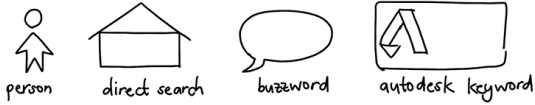

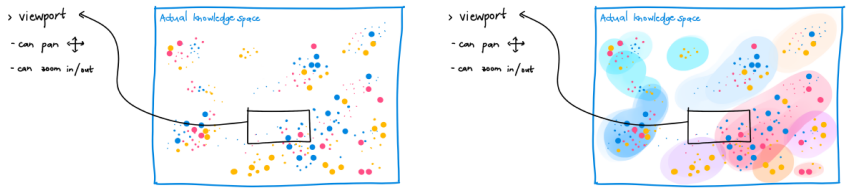

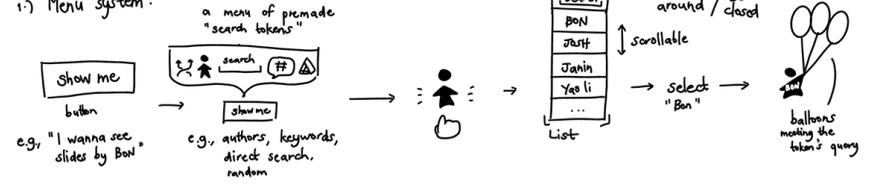

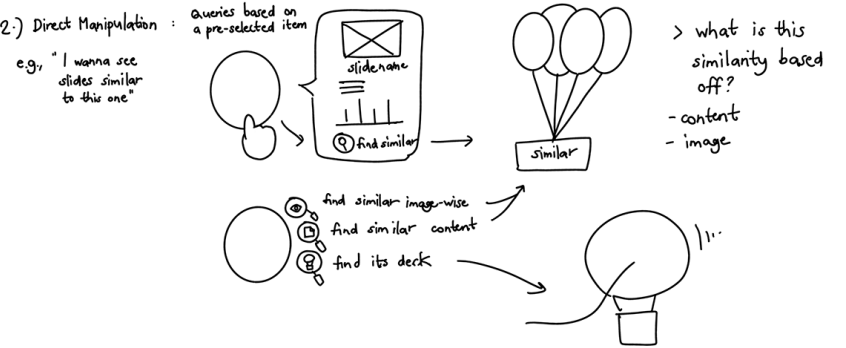

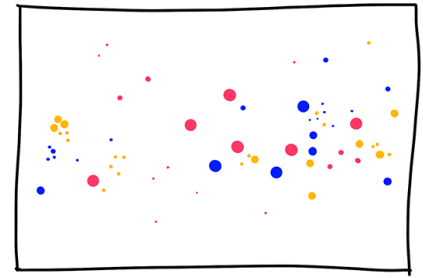

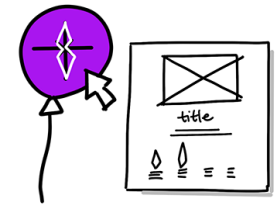

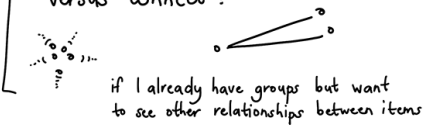

## SkyGlyphs

Following our design sketches, we implemented SkyGlyphs using HTML5/JavaScript, D3 [10], and JavaScript Canvas library. SkyGlyphs uses the *balloon metaphor* to visualize presentation documents (slide decks) that employees shared within the company. We designed SkyGlyphs as an ambient visualization on a large interactive display located in a social area of the company office. The balloons in the datavis are animated to appear as if they are floating in the sky, and their initial position is randomized using D3's force-directed layout that we modified to work on raster images. Balloons that appear closer and large are newer slide decks, while balloons that appear further in the distance (faded and small) are older.

The visualization space of SkyGlyphs could extend beyond what is visible on the screen. To handle this, we implemented an **overview minimap** available on the top right corner of the visualization. This provides a zoomed-out version of the visualization space, including a box that shows where the current view is. A person could traverse the visualization space by touching on the white space and dragging it to the direction they wish to move to. Each balloon could also be dragged, reposition, and pinned.

Much like an ambient art installation, SkyGlyph's menus remain hidden until a person interacts with it. The menus disappear again after some time when the datavis is left on idle. An **explore button** is provided at the bottom of the visualization which displays the **explore dock** when tapped. The explore dock is a menu that contains several options for exploring the visualization (for example, by looking for documents containing certain products and keywords).

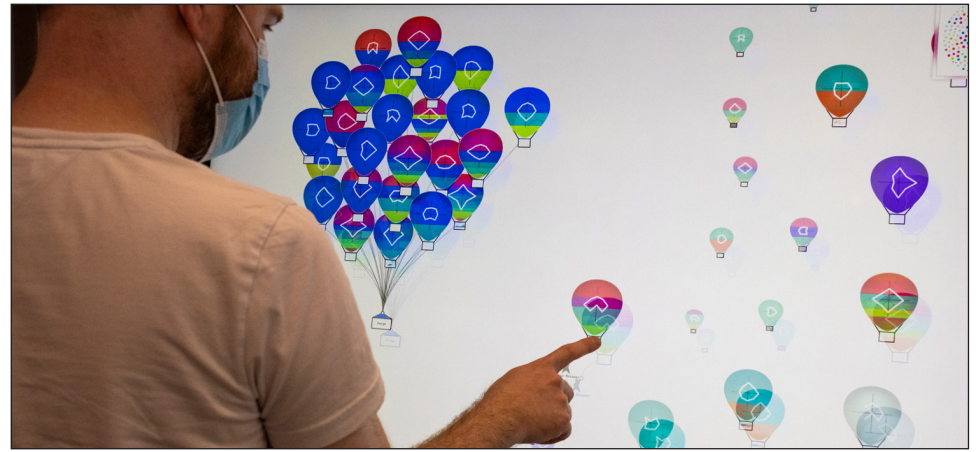
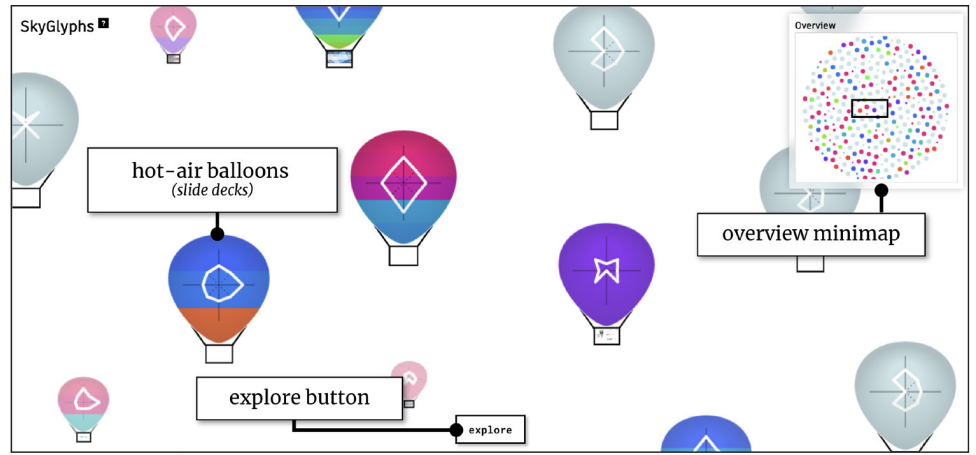
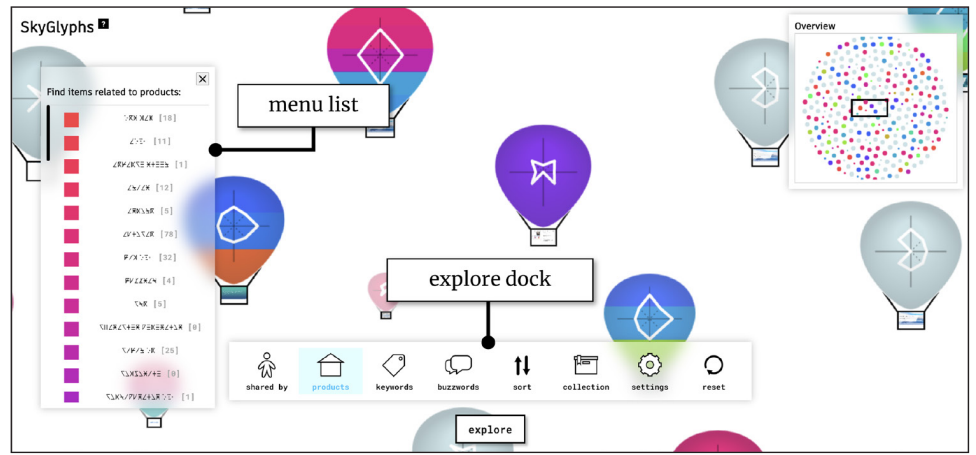

### Balloons

Each presentation document is represented by a **hot-air balloon**. A person could examine a hot-air balloon's slide deck using its **interactive tooltip**. In this tooltip, the **slide preview** shows images of the individual slides contained inside the slide deck. The slide shown in this preview could be changed by interacting with the **slide scrubber** and the **slide thumbnails**. It also contains the slide deck's metadata such as its title, the person who shared it, the repository it was shared in, and the date it was shared. Lastly, the lists of products, company relevant keywords, and common buzzwords mentioned in the slide deck are listed in the tooltip as interactive (clickable) buttons that can be used to find other similar slide decks.

The **spiked glyph** in the middle of the hot-air balloon could be thought of as a radial visualization—a modified *star plot*—with four data axes. To aid its readability, we provided the broken-down spikes next to their corresponding values in the tooltip. Hovering/tapping over these parts will highlight where they are in the hot-air balloon and vice-versa.

Individual slides are represented by smaller, *simple balloons*. To see the simple balloons, a person could long-press on a hot-air balloon to trigger the *balloon popping sequence*. Once a hot-air balloon pops, its inner slides will be displayed as simple balloons that are bundled together. Simple balloons also have an interactive tooltip visualizing its content.

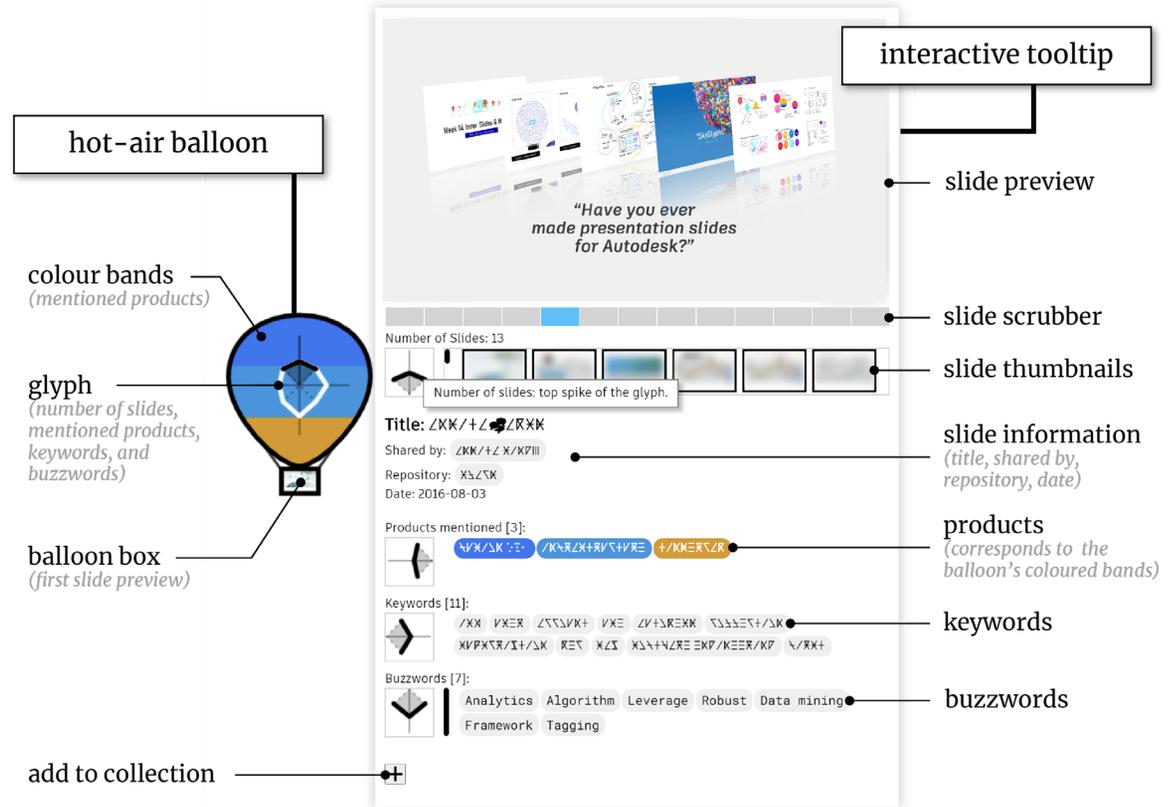

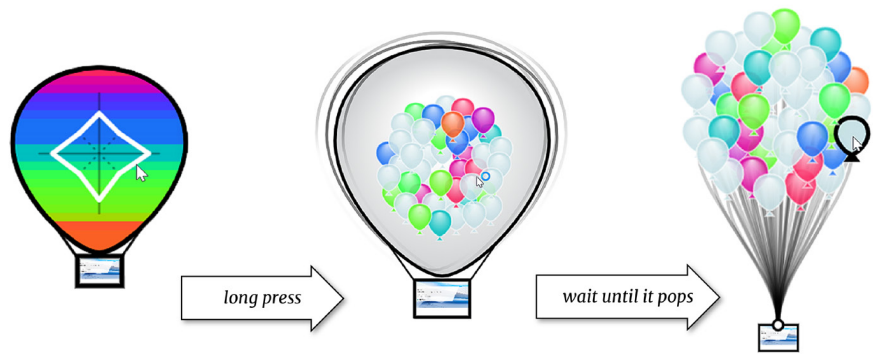

**Balloon popping sequence:** First, the hot air balloon expands, revealing a preview of its inner slides. If the person stops the long-press at this moment, the hot air balloon will revert to its original state, otherwise, the hot air balloon will start shaking until it pops.

**Design of the Spiked Glyph**

We sketched several designs for the glyphs and decided to build upon star plots [14] to design our spiked glyph. Our choice was based on aesthetic considerations such as which style could generate interesting shapes and not collapse to a single point or line for readability, as well as which style could depict numeric data in different scales, while not obviously looking like a regular datavis. We also wanted to ensure it blends well with our balloon metaphor's overall visual.

Typically, star plots act like parallel coordinates with axes (dimensions) radiating from a centre point. The connecting lines are then drawn to the appropriate points in each axis and back, forming a shape. Because of this, star plots are used to view multivariate data in a compact way, usually as shape comparisons [14]. However, if all the values are zero (the centre point), then the shape will collapse to a dot. To prevent this, we modified the star plot to have shorter axes, or *anchor points*, that are always slightly removed from the centre. Furthermore, we used a logarithmic scale on each axis to handle large discrepancies between some data-attribute values. For instance, we found slide decks that contain hundreds of slides, while only having less than 10 keywords. In this case, the spike corresponding to the number of slides will exceed the allotted glyph space when using a linear scale.

While this adds visual complexity, it could form more interesting shapes. After all, our intention for the spiked glyphs is **not** so people could use it to accurately analyze the numeric data. Rather, we envisioned them to act as *visual hooks* to make people curious. This is in line with research in psychology stating that *complexity* and *uncertainty* are motivators for curiosity [8,20]. A person could become curious when they encounter something that they cannot explain, making them want to learn about it.

Glyphs

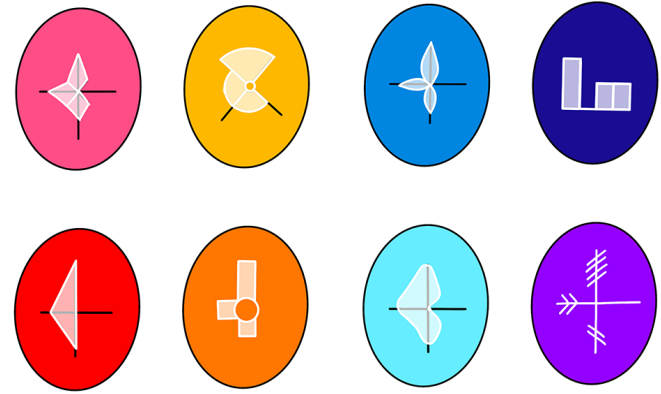

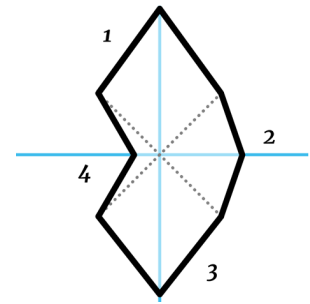

The spiked glyph could be thought of as radial visualization (star plot) with four axes (shown in blue). The shorter axes (shown as dotted lines) act as anchors to create the spikes.

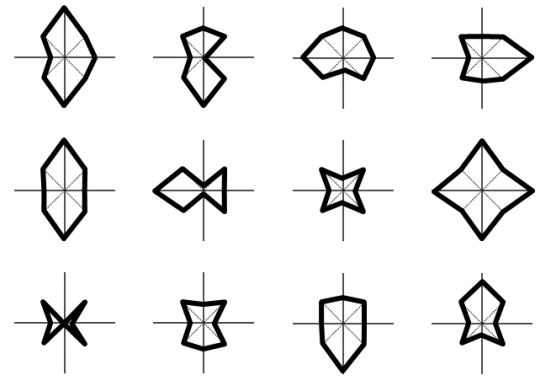

The anchored glyph design allows for interesting shapes to form compared to a typical four-axis star plot that could collapse to a single line or a dot.

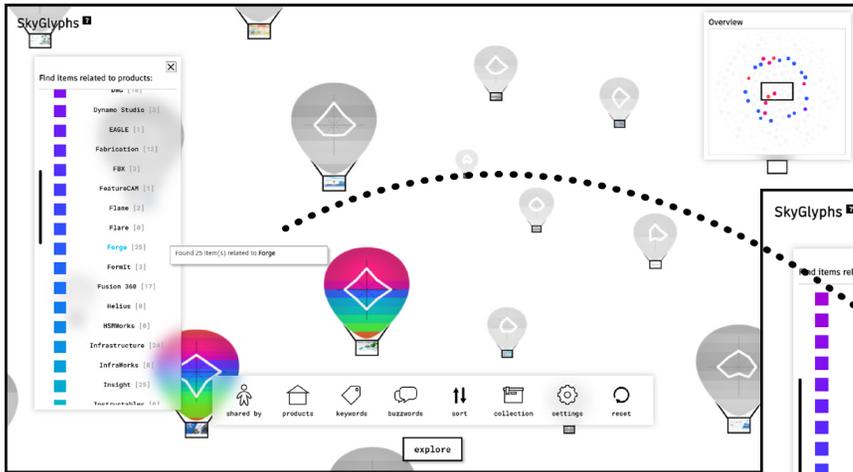

**Exploring the Visualization: Clusters**
The explore dock menus could be used to create clusters of balloons based on a shared property. For example, one could cluster slide decks that mention the same product name.

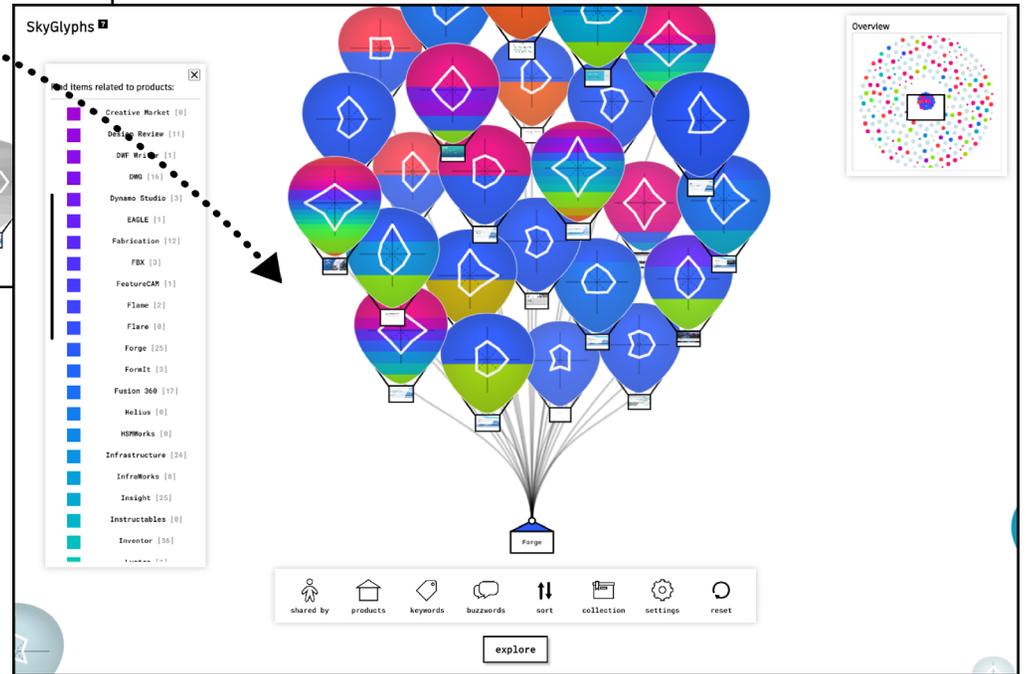

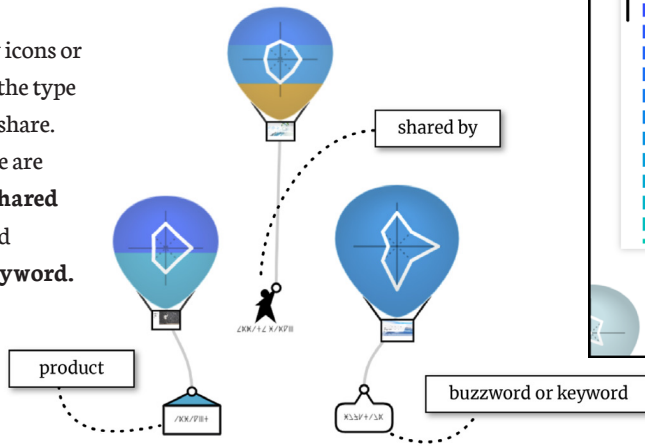

Clusters are differentiated by icons or *anchors* based on the type of property they share. For example, here are the anchors for **shared by, product,** and **buzzword** or **keyword.**

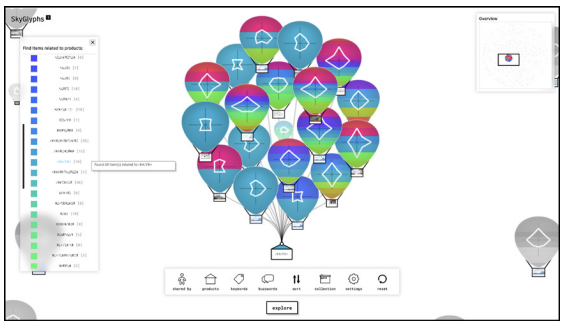

A single balloon cluster based on a mentioned product.

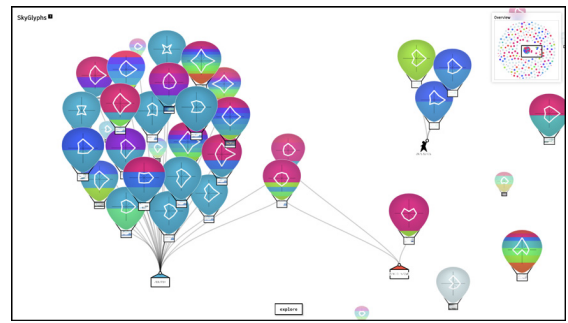

Two clusters with shared balloons.

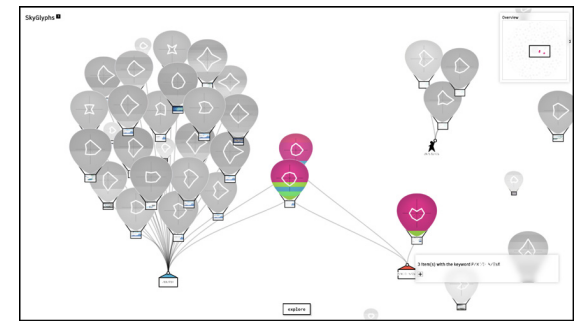

Hovering over a cluster's anchor will highlight the balloons that belong to it.

VISAP'22, Pictorials and annotated portfolios.

**Exploring the Visualization: Overview**

Depending on the screen size, SkyGlyphs could only show a portion of the balloons on screen at a time. To see more datapoints, a person could zoom-out of the visualization to its **overview** that shows most if not all the datapoints at once (see below). In this view, the hot-air balloons are simplified into *dots*. These dots are small and thus could no longer accommodate details like the spiked glyph. They are also further simplified by having only one colour based on the dominant product mentioned in its slides.

**Exploring the Visualization: Sorting**

The explore dock also allows a person to sort and rearrange the balloons based on a data attribute in ascending or descending order. When sorted, the balloons are listed to fill the available screen space, first from left to right, and then top to bottom. Only the vertical space can have an overflow for items that do not fit the screen space. For example, at the bottom of the figure below, six hot air balloons are sorted in descending order according to the number of slides they contain. This is also reflected on the balloons' spiked glyph where the top spike, representing the number of slides, successively gets shorter in each balloon.

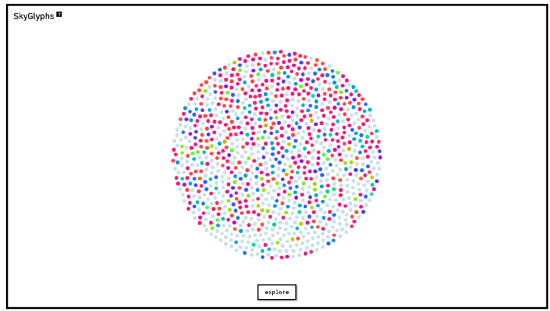

zooming out into the *overview* will show all of the data–points as dots

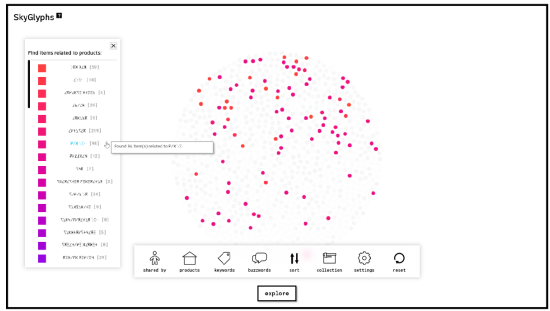

the exploratory interactions are still available in the overview

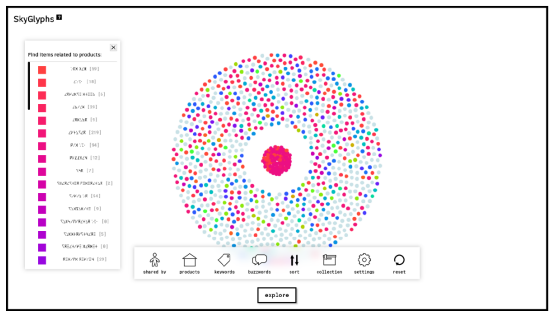

any layout changes committed while in the overview will persist when reverting to the default view

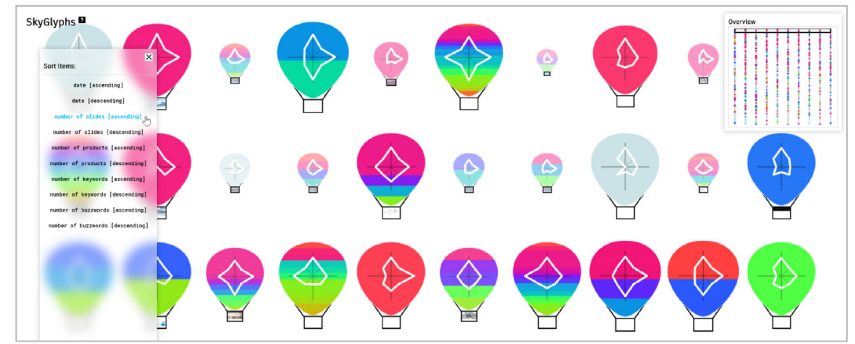

*example sort:* ▼ *number of slides*

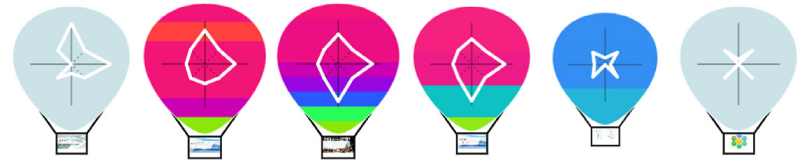

## Collections

A person could add items of interest to the **collections panel** (image on the right) by clicking the **+ add to collection button** (a) in the items' tooltip, or by dragging the items into the panel (b). These items could be individual balloons or clusters of balloons. To explore their collection, a person could toggle a filter for showing only the collected items using the commands at the bottom of the collections panel. These commands include *filter* or hiding and unhiding items that are not in the collection, and *clear*, which removes all the items in the collection.

## In Perspective

We, the authors, come from different backgrounds in HCI, Infovis, Art and Design, and Industry. As we provide our reflection, we would like to acknowledge that our process of designing and implementing a delightful data visualization was shaped by our own perspectives, experiences, and limitations, alongside typical HCI methods like understanding our end-users' contexts and tasks to support. However, it is in this *mingling of perspectives* where we find insights worth sharing to the visualization community.

## Visual Metaphors and Themes

We carefully crafted SkyGlyphs' aesthetic design by first selecting an inspiration as a visual metaphor to represent datapoints in our datavis—*balloons*. In doing so, we were able to formalize the visual metaphor using intrinsic and extrinsic visual qualities of balloons—their shapes, colours, patterns, and behaviours. We also considered the inherent *fun* associated with balloons being associated with positive ideas like celebration and freedom. These qualities could be leveraged to direct a visualization's *look-and-feel*. In our design process, we took note of these characteristics and studied how to incorporate them in the visualization. We interpret this step in our design process as formalizing *visual metaphors* and *themes*. The metaphor is the balloon and how it could represent the data attributes, while the theme involves the overall experience and behaviour of "balloons floating in the sky."

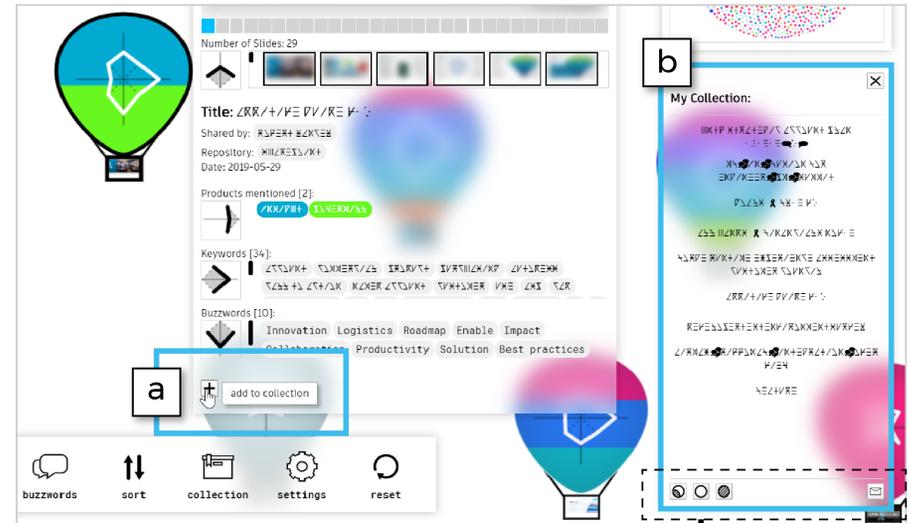
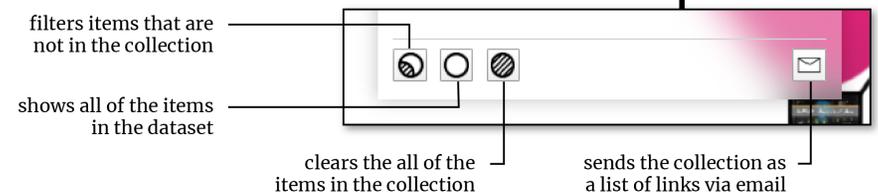
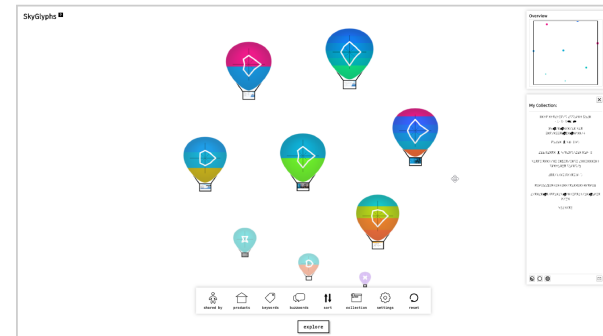

The collections panel can be used to collect items a person is interested in. It can also be used to focus on only a few items at a time

## Inspiration

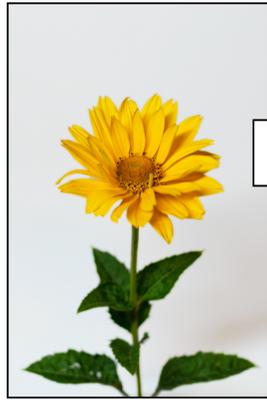

*flower*

## Visual Metaphor

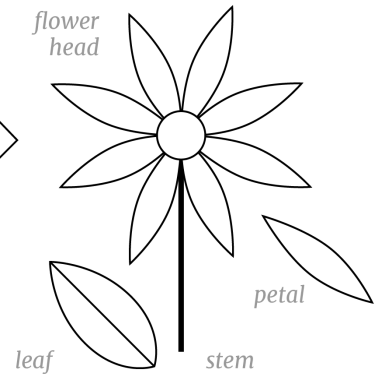

*flower head*

*petal*

*leaf*   *stem*

*essential forms*

→ *abstraction* →

*dataset*
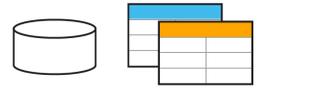

→ *visual encoding* → *data mapping* →

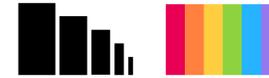
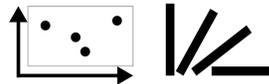

*visual variables*

## Data Representation

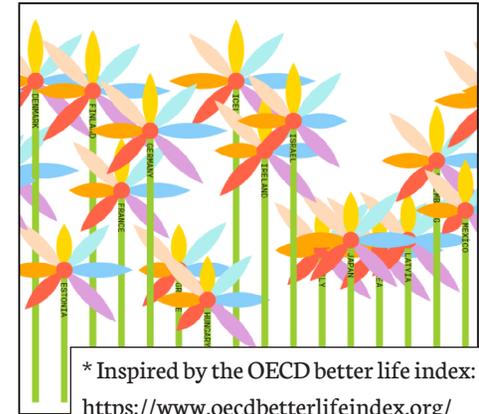

\* Inspired by the OECD better life index: https://www.oecdbetterlifeindex.org/

This design process phase is an exercise in *visual abstraction*, or the process of reducing a complex object into its set of essential visual characteristics. For example, a flower can be broken down into basic shapes that resembles its flower head, petals, stem, and perhaps, leaves. This then becomes a visual metaphor for a datapoint in a dataset. The datapoint's attributes can then be mapped into these essential forms through their visual channels or variables like colour, length and size, position, and orientation. The theme then dictates how the flowers would look and feel like. Depending on the mood of the theme, the flower visual metaphor could be a rose, or a lily, all of which depend on aesthetic choices. Combining all the metaphors into a themed view results in a data representation. In finalizing the design solution, it may be important to consider the cultural meaning of the inspirations.

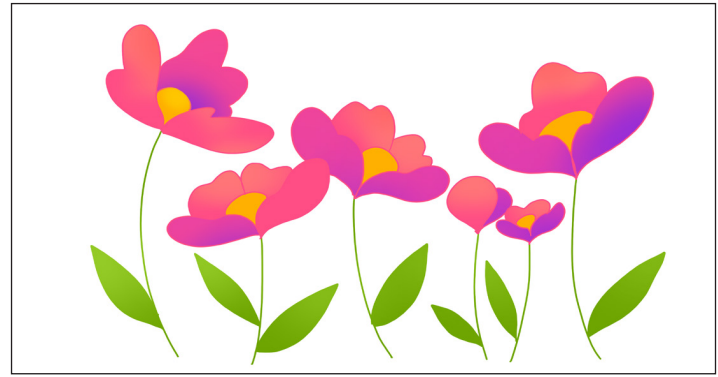

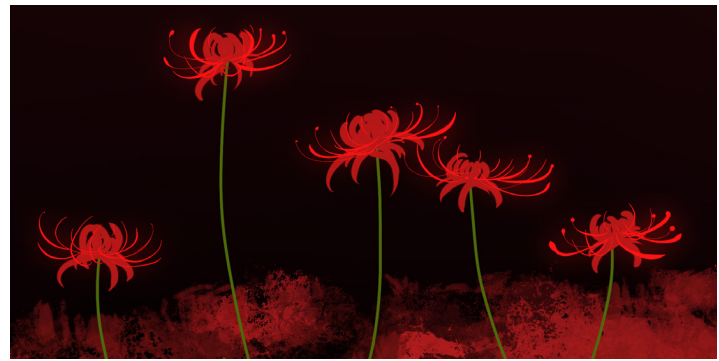

Both of these illustrations use a flower *visual metaphor*, however both have very different look-and-feel (*theme*). These characteristics can be intentionally used when designing nonconventional datavis that appeal to different emotions.

However, we acknowledge that some of the abstractions we implemented do not necessarily depict real-life behaviour of our chosen metaphor. For instance, while normal balloons could *pop*, hot-air balloons in real-life (ideally) should not, nor do they contain internal balloons. We designed this feature in SkyGlyphs to bridge our whimsical metaphor with common set visualization like circle packing, and interactions like gradual reveal (or details on demand).

We do not claim that this is unique to our process but rather an interpretation from our reflection. We think of this as a means to help articulate authorial intent in visual abstraction. For instance, one may apply this method to examine how other data visualizations used balloons as a metaphor [16,31], or when finding their own inspirations for designing new metaphors. Our process mirrors that of Cruz's "semantic figurative" approach in which embellishments are used to make an emotional connection with the audience [13].

### Whimsical Aesthetics
Whimsical aesthetics could be used to create fun and invitingly themed visualizations. To make SkyGlyphs delightful, we considered aesthetic choices such as using vibrant colours, elegant shapes, and fun, satisfying animations. We also deliberately chose a metaphor and theme that is playful and amusing, which we consider as *whimsical*. Visual embellishments have previously been thought to be detrimental to the success of a visualization. However, the Infovis community now generally agree that carefully designed embellishments can help visualizations become more memorable, helping people recall both the visualization and the data it represents [7,9].

One of our main challenges in designing SkyGlyphs was to encourage employees to explore presentation documents in the hopes that they might be able to find interesting information that could be beneficial to them and by extension, the company. If SkyGlyphs' design enables people not just to browse presentation documents but also helps them remember what they have explored, then that is definitely advantageous. This could be part of a future investigation on the possible advantages of nonconventional visual representations in emergent contexts. On a similar note, it will also be interesting to study how far whimsical and/or delightful aesthetics can go before they become uninviting or even inappropriate for certain contexts.

### Unit Visualizations and Glyphs
SkyGlyphs represents each datapoint as a "point of interest," thus functioning as a *unit visualization* [22]. According to Park et al., people can easily interpret unit visualizations because of their intuitive one-to-one mapping between datapoints and visual marks. Unit visualizations also conform with constructivist notions in Infovis, or the idea that novice datavis users can understand and construct visualizations by representing datapoints as individual tokens [18]. While this requires further investigation, this could mean that unit visualizations like SkyGlyphs are primed for use by the general population (like non-experts), making them ideal for personal visualizations.

Glyphs are graphical entities that represent a single, multivariate datapoint [29], thus, they are also a type of unit visualization. Each datapoint in SkyGlyphs is a complex glyph. For instance, a single hot-air balloon depicts a variety of data attributes of a slide deck, and most of them look unique, differing slightly in the data attributes they express. This hints that complex glyphs have the potential to surface a lot more information about a datapoint and be more expressive. Nevertheless, this visual complexity can make the visualization harder to read. Future work could examine the trade-offs between glyph complexity and expressiveness. For visualizations like SkyGlyphs, a study looking into the dynamics between the glyphs functioning as curiosity-inducing hooks and people's ability to read and interpret them is a good area to begin.

### Bridging Artistic and Analytical Visualizations
SkyGlyphs has potential to act as a bridge between nonconventional visualizations, and more analytical visualizations. We see this from SkyGlyph's overview feature in which a person zooms out to see the whole dataset, transitioning to a more scatterplot-type visualization without embellishments. In the overview, people could reconfigure the visualization using the exploratory features like sorting, grouping, and such. Reflecting on this aspect, we think that the overview feature could also enable people to do more analytical tasks. While in the overview, we envision SkyGlyphs to provide an *analytics tool palette* containing tools for analysis inspired by SandDance [21]. For example, an **add axis tool** could let people draw axes on the visualization space, creating a scatterplot view for exploring the dots. This could enable some people to do a deeper analysis of the datapoints without needing to move to a different visualization platform. This could be part of future work exploring what delightful visualizations could support once they have captivated people's attention.

## Conclusion

In this pictorial, we presented our design process in creating a SkyGlyphs, a nonconventional whimsical visualization that enables free-form exploration of a dataset. SkyGlyphs' visual metaphor and theme is inspired by balloons floating in the sky. Through this delightful and possibly memorable aesthetic, SkyGlyphs' leverages people's *curiosity* as an entry point to data exploration. We also present the final implementation of SkyGlyphs and illustrated its interactive features.

Overall, our process in designing SkyGlyphs opens avenues for future studies about nonconventional visualizations. This includes investigating whether whimsical aesthetics could truly make visualizations more curious, memorable, and inviting to use. Our hope is that it offers a new perspective and insights as to how one might design new nonconventional data visualizations.

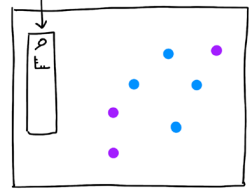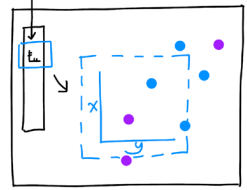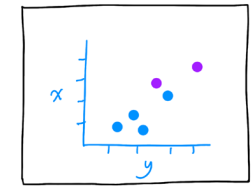

## Acknowledgement

Some ideas in this pictorial such as *visual metaphors, themes, and whimsical aesthetics* are explored in more detail in the first author's thesis [3].